# Environmental Pollution Prediction of NOx by Process Analysis and Predictive Modelling in Natural Gas Turbine Power Plants

Alan Rezazadeh, Southern Alberta Institute of Technology
Calgary, Alberta, Canada

# 1 Abstract

The main objective of this paper is to propose K-Nearest-Neighbor (KNN) algorithm for predicting NOx emissions from natural gas electrical generation turbines. The process of producing electricity is dynamic and rapidly changing due to many factors such as weather and electrical grid requirements. Gas turbine equipment are also a dynamic part of the electricity generation since the equipment characteristics and thermodynamics behavior change as the turbines age. Regular maintenance of turbines are also another dynamic part of the electrical generation process, affecting the performance of equipment. This analysis discovered using KNN, trained on relatively small dataset produces the most accurate prediction rates. This statement can be logically explained as KNN finds the K nearest neighbor to the current input parameters and estimates a rated average of historically similar observations as prediction.

This paper incorporates ambient weather conditions, electrical output as well as turbine performance factors to build a machine learning model to predict NOx emissions. The model can be used to optimize the operational processes for reduction in harmful emissions and increasing overall operational efficiency. Latent algorithms such as Principle Component Algorithms (PCA) have been used for monitoring the equipment performance behavior change which deeply influences process paraments and consequently determines NOx emissions. Typical statistical methods of machine learning performance evaluations such as multivariate analysis, clustering and residual analysis have been used throughout the paper.

# 2 Introduction

The main objective of this paper is introducing K-Nearest-Neighbor algorithm as a candidate to be used in Predictive Emission Monitoring Systems (PEMS), predicting Nitrogen Oxides (NOx) emissions produced in the process of electricity production of gas turbines. This paper used the gas turbine process parameters dataset from University of California at Irvine (UCI) open data repository [1], which was collected over five year period in north western Turkey. The power generation utility donated the dataset would like to remain anonymous and author would like to extend gratitude for allowing this valuable dataset to be used in research. The power plant location is close to sea level, prone to humidity fluctuations, comprised of mild temperatures occasionally dropping below freezing point.

The power generation system is a Combined Cycle Power Plant (CCPP), comprised of gas and steam turbines. Figure 1, depicts schematics of the power plant, comprised of two gas turbines of 160MWh each with a Heat Recovery Steam Generator (HRSG) powering a 160MW steam turbine. The exhaust from gas turbines, usually maintain high temperatures are used for driving a steam turbine, which result in a highly efficient power generation system, approximately about 60% efficiency in comparison to a simple cycle gas turbine of approximately 35% to 40% efficiency [7].

PEMS have been discussed widely within literature for the last 25 years as a backup to Continuous Emission Monitoring Systems (CEMS) using allocated sensors, directly measuring emissions and pollutants produced by the combustion process [15]. CEMS based on dedicated hardware and necessary software have been a part of gas turbine design with well-defined legal operational requirements. Many researchers consider PEMS as a backup, or alternate monitoring system to CEMS. This paper presents another application of PEMS, which is monitoring the electrical generation process efficiency, in addition to predicting emissions such as NOx under fast changing process conditions.

As the field of industrial data science is evolving, more applications to PEMS are being identified. A new application to use PEMS can be identified as monitoring degradation and process efficiency of gas turbines. PEMS, naturally using many process parameters such as turbine pressures and temperatures for predicting emissions, which are in great position to monitor the process performance and degradation.

PEMS as a method of predicting emissions uses operational data for training and building machine learning models. As the training data becomes longer in time, the electrical generation process may change due to new grid requirements, extreme weather conditions or equipment may have different performance due to degradation, hence decreasing the prediction success. As a result, contrary to popular believe, shorter training time may actually offer better prediction rates, utilizing more of the recent data points, rather than including all historical data.

The paper discusses application of latent variables such as principle components for monitoring and early detection of process change. Benchmarking process behavior based on physics of gas turbines and laws of thermodynamics also present another method of monitoring process drift or degradation. However, due to lack of sufficient internal turbine data this paper refrains from exploring the degradation in more depth and invites the power generation industry to share more detailed internal process dataset for further analysis and research.

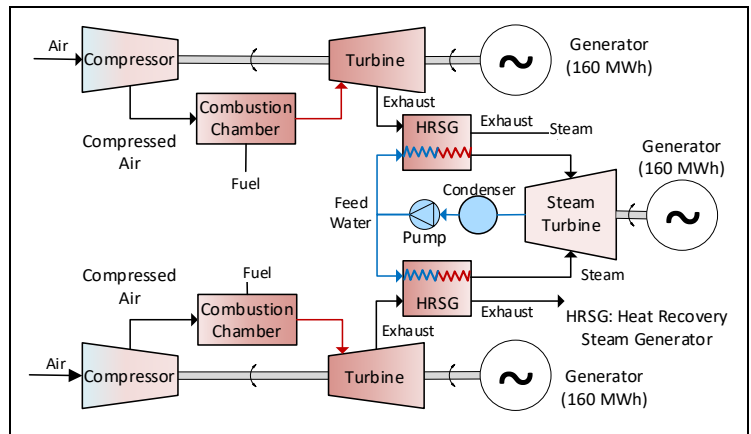

*Figure 1 Combined cycle power plant schematic diagram*

## 3   Gas Turbine Emissions

NOx is a generic term for emission family of Nitrogen Oxide (NO2) and Nitric Oxide (NO), which are usually created as a result of combustion process. Although, both transportation and power generation sectors use combustion process, the main objective of this paper is NOx resulted from gas power plants, which contribute to smog, acid rain and tropospheric ozone [14].

Even though, natural gas is relatively a cleaner burning fossil fuel, combustion process produces small amounts of sulfur, mercury and particulates depending on the quality of fuel. These pollutants are considered fuel dependent and can be eliminated by using higher quality and cleaner natural gas. In contrary NOx considered pollutant which is resulted from higher temperature combustion such as gas turbines. NOx are considered process dependent meaning by optimizing combustion process the pollutant can be minimized. Table 1 illustrates typical pollutant emissions from gas turbines. The main objective of this research is to better understand and predict NOx resulted from combustion process in gas turbines.

| Gas Turbine Pollution | | |
|---|---|---|
| Pollutant | Fuel Dependent | Process Dependent |
| NOx |  | ✓ |
| CO |  | ✓ |
| SOx | ✓ |  |

*Table 1. Typical pollution emissions from gas turbines and their source*

## 4 Predictive Emission Monitoring Systems

Predictive Emission Monitoring Systems (PEMS) are software solutions for predicting emissions based on operating process parameters such as internal turbine pressures and temperatures. Emissions resulted from combustion process are usually monitored and measured using CEMS; hardware sensors in two different methods of periodical or continuous intervals. In either case, specialized hardware is used and usually maintenance of the sensors are within the project budgets and schedules.

Figure 2 illustrates the schematics of an open cycle gas turbine model [10] and the data elements available for this study. The existing data elements describing the gas turbine operations, miss a few critical elements for a more thorough analysis such as fuel amount, compressor discharge temperature and released Carbon Dioxide. Nevertheless, existing data offers valuables insights into operations of gas turbines and predictions of NOx. The available data elements can be found in Figure 2 as well as Table 2.

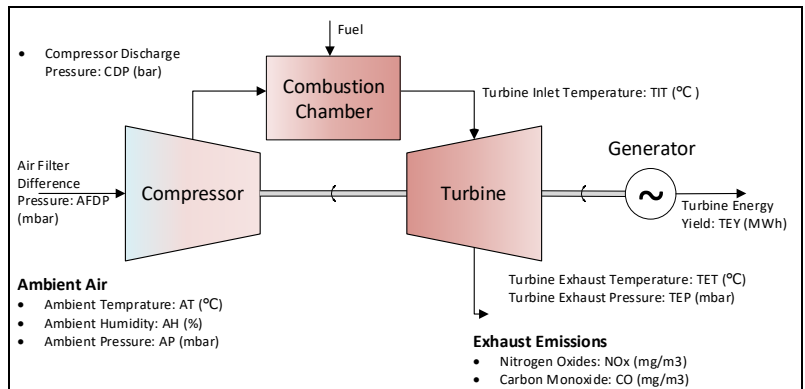

*Figure 2. Schematic diagram of a simple cycle gas turbine, including the process parameters used in the study*

## 5 Data Exploration

Industrial data analytics begin with understanding the dataset including their internal relationships, trends and distributions. In most industrial applications the dataset may contain tens of thousands of data points (i.e. records) with tens of variables (i.e. predictors) or more. The variables are typically sensor readings within a process and may contain strong collinearly, meaning groups of predictors may move together under specific conditions.

### 5.1 Univariate Analysis of Process Parameters

Table 2, illustrates the histogram of power generation process parameters for the period of five years, 2011 to 2015, beginning from January 1st of each year. As can be seen there are overall 12 variables, including six internal variables to the gas turbine, which are used to describe the status of power generation process.

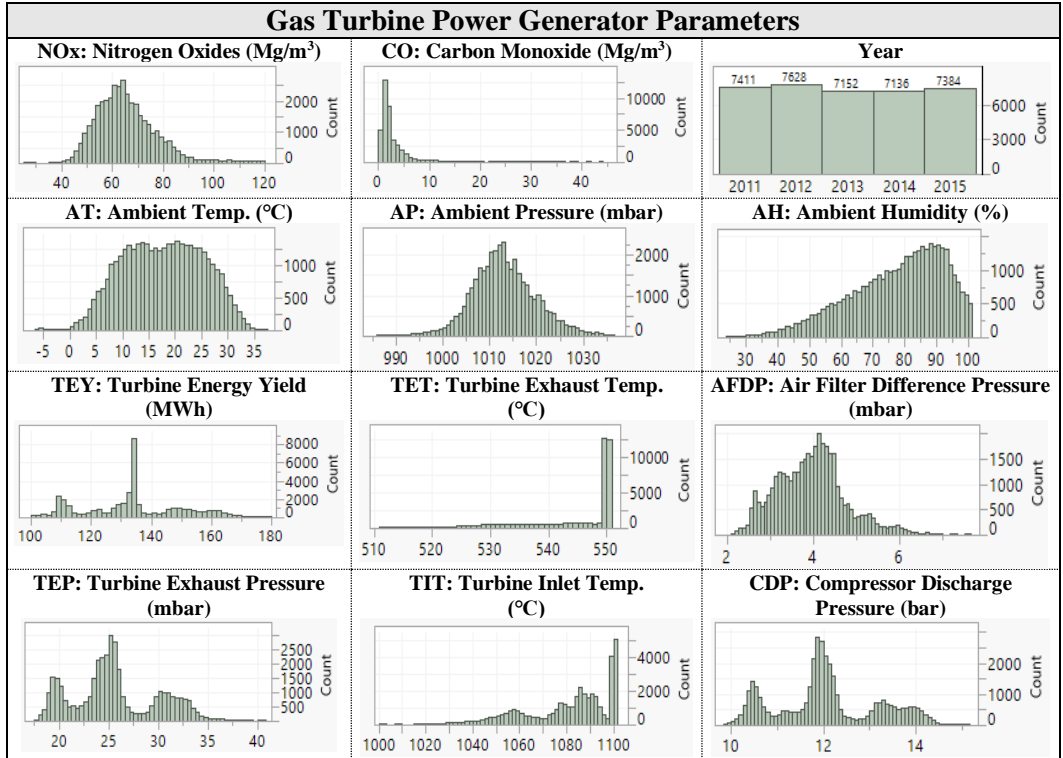

Table 2. Gas turbine power generator variables

## 5.2 Multivariate analysis of Process Parameters

Multivariate analysis is a set of techniques to analyze the multidimensional data, seeking patterns within data elements. Table 3 illustrates the multivariable correlation between the variables in two formats of numerical correlation and visual scatter plot. The red data points in the scatter plots indicate data readings with higher NOx values.

As can be seen in Table 3, the higher values of NOx are usually clustered closer to lower temperatures and lower power production yields, which result lower turbine temperatures and pressures. Visual inspection of scatter plots in Table 3, and correlation values indicate the process variables are more correlated with each other and less with the weather condition parameters. Therefore, clustering parameters in a scientific and quantitative method can clarify the relationships in more details.

|  | AT (C) | AH (%) | AP (mbar) | TIT (C) | TET (C) | TEP (mbar) | AFDP (mbar) | CDP (bar) | TEY (MWh) |
|---|---|---|---|---|---|---|---|---|---|
| AT (C) | 1.0000 | -0.4763 | -0.4067 | 0.1840 | 0.2821 | 0.0458 | 0.2519 | 0.0152 | -0.0914 |
| AH (%) | | 1.0000 | -0.0153 | -0.2218 | 0.0235 | -0.2350 | -0.1478 | -0.1961 | -0.1371 |
| AP (mbar) | | | 1.0000 | -0.0043 | -0.2252 | 0.0580 | -0.0403 | 0.1031 | 0.1190 |
| TIT (C) | | | | 1.0000 | -0.3867 | 0.8750 | 0.6928 | 0.9093 | 0.9106 |
| TET (C) | | | | | 1.0000 | -0.7026 | -0.4680 | -0.7094 | -0.6861 |

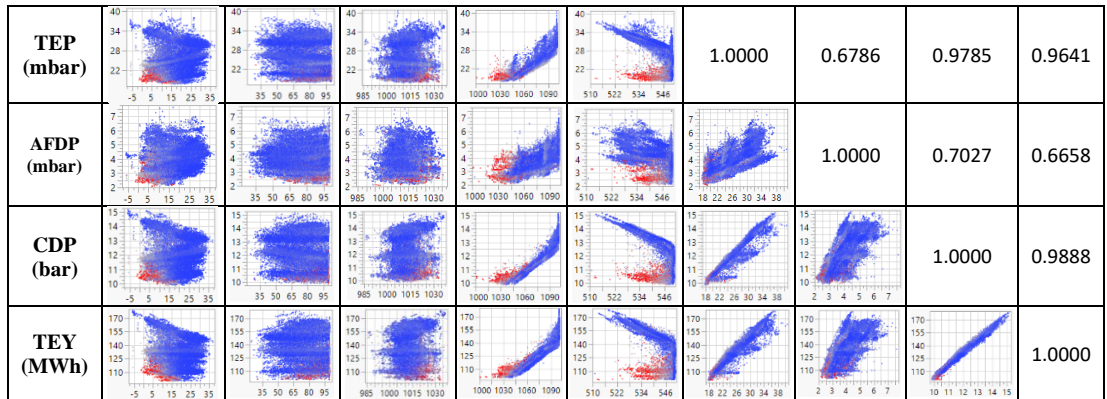

Table 3. Multivariate analysis of available predictors. Red data points indicate higher NOx values.

## 5.3 Clustering of Variables

Clustering of variables (Table 4), is performed by using principle components, based on application of eigenvalues and eigenvectors. At the beginning, all variables are assigned to one cluster, if the second eigenvalue is larger than a predefined threshold, the variables split into two clusters, since second large eigenvalue means variations among the second group of variables. The process continues until the second eigenvalues of all clusters fall below the predefined threshold [9].

Table 4 shows the given variables can be split into three clusters. Interestingly as can be seen all members of cluster 1, are the internal parameters of gas turbine, meaning the variables are highly correlated. The second cluster consists of Ambient Humidity (AH) and Ambient Temperature (AT), meaning these two variables are correlated and move together. The third cluster consists only of one variable which is Ambient Pressure (AP), meaning this variable is independent of other factors.

| Cluster | Members | RSquared with Own Cluster | RSquared with Next Cluster | 1 – RSquare Ratio | Comments |
|---|---|---|---|---|---|
| 1 | CDP (bar) | 0.983 | 0.015 | 0.017 | Process Dependent |
| 1 | TEY (MWH) | 0.959 | 0.014 | 0.041 | Process Dependent |
| 1 | TEP (mbar) | 0.951 | 0.027 | 0.050 | Process Dependent |
| 1 | TIT (C) | 0.816 | 0.056 | 0.195 | Process Dependent |
| 1 | AFDP (mbar) | 0.602 | 0.054 | 0.421 | Process Dependent |
| 1 | TAT (C) | 0.523 | 0.051 | 0.503 | Process Dependent |
| 2 | AH (%) | 0.738 | 0.034 | 0.271 | Weather Dependent |
| 2 | AT (C) | 0.738 | 0.165 | 0.314 | Weather Dependent |
| 3 | AP (mbar) | 1.000 | 0.052 | 0.000 | Weather Dependent |

Table 4. Clustering of process parameters

For the first cluster, Compressor Discharge Pressure (CDP) contains 98.3% of the variation within the cluster. Meaning using CDP is the best variable of this cluster (gas turbine parameters) to explain the variance. Interestingly CDP is one of the most important factors in predicting efficiency of gas turbine Brayton cycle thermodynamics [10] and also electrical yield has very strong linear relation with CDP.

R-Squared for the cluster variables are defined as the ratio of unexplained variance on a variable to its own cluster component [9]. R-Squared with next cluster is the proportion of explained variance within a variable with the next cluster. The value of 1-RSquared ratio is defined as the ratio of 1 minus its own cluster R-Squared to 1 minus next closest cluster's R-Squared.

$$1 - RSquared = \frac{1 - RSquared\ with\ own\ Cluster}{1 - RSquared\ with\ Next\ Closest}$$

*Equation 1. Definition of 1 – RSquared Ratio*

Based on Table 4, there are three clusters identified, which CDP, AH and AP are the most significant parameters with the most variance for each group.

### 5.4 Predictor Screening based on Process Parameters

Predictor screening is used to find the contribution of each predictor to the response variable, NOx values. This technique is specifically advantageous for variables with potentially weak direct correlation with the response (NOx); however, with stronger interaction through other variables. Predictor screening is based on Bootstrap Forest [16] (fitting a model by averaging many trees similar to Random Forest), finding contribution and percentage portion of contribution to NOx values. Table 5 shows the most contribution was made by AT (Ambient Temperature) to NOx production, with overall 31.8% of all effects.

| Predictor Screening - NOX (mg/m3) (Process and Weather Parameters) | | | | |
|---|---|---|---|---|
| Predictor | Contribution | Portion | | Rank |
| AT (C) | 631626 | 0.3185 | | 1 |
| TIT (C) | 305719 | 0.1541 | | 2 |
| TEP (mbar) | 221869 | 0.1119 | | 3 |
| TET (C) | 205984 | 0.1039 | | 4 |
| AFDP (mbar) | 181085 | 0.0913 | | 5 |
| TEY (MWH) | 173212 | 0.0873 | | 6 |
| CDP (bar) | 136176 | 0.0687 | | 7 |
| AP (mbar) | 93584 | 0.0472 | | 8 |
| AH (%) | 34009 | 0.0171 | | 9 |

*Table 5. Identification of most related predictor to NOx production, including weather parameters*

Excluding the weather parameters from the predictor screening and running the analysis with only turbine internal process predictors, results Table 6. As can be seen the order of parameters are still similar to Table 5 which included weather parameters as well; however, only the contributions to NOx production are different.

| Predictor Screening - NOX (mg/m3) (Only Process Parameters) | | | | |
|---|---|---|---|---|
| Predictor | Contribution | Portion | | Rank |
| TIT (C) | 234449 | 0.2205 | | 1 |
| TEP (mbar) | 202488 | 0.1904 | | 2 |
| AFDP (mbar) | 174308 | 0.1639 | | 3 |
| TET (C) | 172821 | 0.1625 | | 4 |
| TEY (MWh) | 156293 | 0.1470 | | 5 |
| CDP (bar) | 122904 | 0.1156 | | 6 |

*Table 6. Identification of most related predictor to NOx production, without weather parameters*

The predictor screening results indicate the strongest factor including the weather data is ambient temperature, which may influence consumers for using more power during colder hours. Increased power demand, may force power generation to operate on higher yield modes, increasing TIT and CDP which result in reduction of NOx, and increased NOx production during lower demand hours in combination to colder air intake.

## 6 Operations Analysis

The dataset used in study contains five years of hourly process parameters as described in Table 2. The multivariate analysis of predictors (Table 3) illustrates the aggregated correlation between predictors over five years of study, without including the effects of time. During five years of operations many correlations may change due to variety of factors such as weather (i.e. abnormally low or high temperatures), consumer demand change, grid requirements or equipment degradation, which would require operators to adjust process parameters for most efficient operations.

For a complete analysis the effects of time and process change more detailed dataset is required. While the exact prognosis of process change will be considered out of scope for this paper, which also requires more detailed turbine parameter data and power grid demand.

### 6.1   Process Change Over Time

As illustrated in Table 5, the most contributing factor to NOx production is ambient temperature. The ambient temperature also affects the consumer demand on power which ultimately defines other process parameters such as turbine energy yield, operational pressure and temperature parameters.

Figure 3. Bivariate analysis of turbine power output versus ambient temperature by yearillustrates the total energy yield versus ambient temperature. As can be seen the red data point indicating higher levels of NOx, are consistently at lower temperatures and lower power production range. Reviewing Table 3, TEY versus TIT chart indicates NOx are mostly produced at lower temperatures and lower electric power yields that can be due to lower consumer demand (off peak hours) or operational start up and shut-down periods.

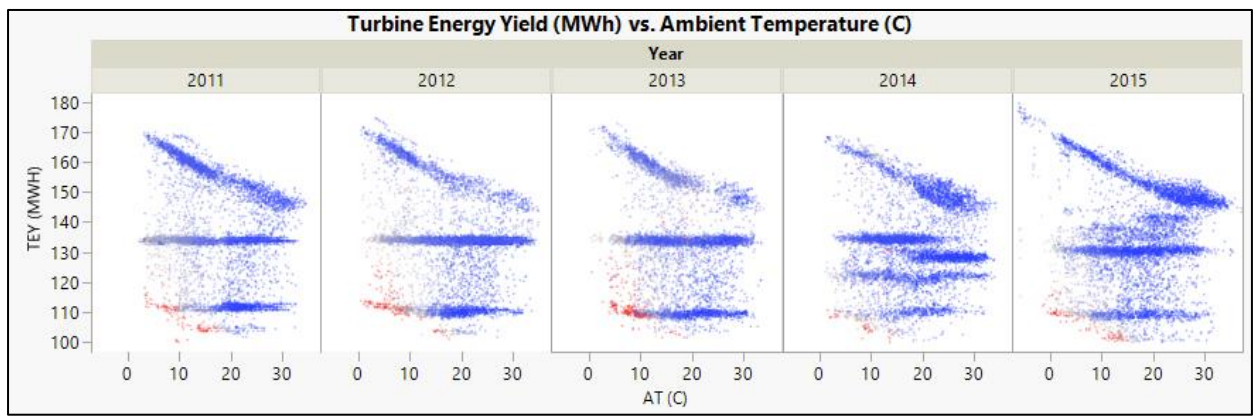

*Figure 3. Bivariate analysis of turbine power output versus ambient temperature by year. Red data points indicate higher values of NOx.*

Analysis of equipment parameters such as CDP (Compressor Discharge Pressure) versus TEP (Turbine Exhaust Pressure) by year depicts the physical characteristics change over time. As can be seen in Table 7, the relationship of CDP versus TEP, which is a hardware parameter and defined by laws of thermodynamics have changed over the life of data set. This change highlights the reason static prediction models will lose accuracy over time as the process is changing; instead adaptive models will be required to be trained only on smaller, more recent data points predicting smaller range into future.

As the process is changing (Table 7), a quantitative benchmark should be used for accurate and unbiased process monitoring [3]. The existing dataset was originally intended for predictive modeling of NOx production, which lacks required data elements for efficiency analysis of the gas turbine. Identifying the exact reasons and root cause analysis of process drift requires more detailed data points, for instance more turbine pressure and temperature readings to cross reference with thermodynamics and gas laws of turbines.

| 2011 | 2012 | 2013 | 2014 | 2015 |
|---|---|---|---|---|
| 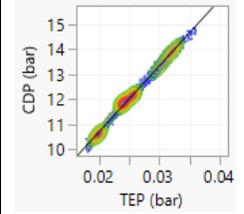 | 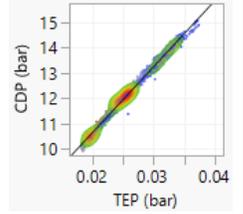 | 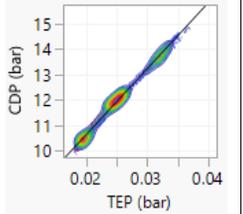 | 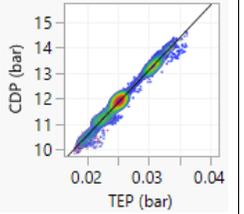 | 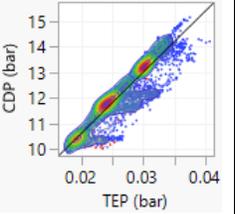 |
| CDP = 5.44 + 263.60 * TEP | CDP = 5.44 + 260.40 * TEP | CDP = 5.46 + 259.22 * TEP | CDP = 5.46 + 254.93 * TEP | CDP = 5.86 + 238.34 * TEP |
| R-Square: 0.9891 | R-Square: 0.9885 | R-Square: 0.9923 | R-Square: 0.9808 | R-Square: 0.8801 |

*Table 7. Process change and degradation over time*

## 6.2 Operational Change Detection by Latent Parameter Analysis

Principle Components Analysis (PCA) is a dimension reduction method to reduce redundancy in a larger set of variables, generating smaller number of orthogonal vectors, preserving the information as much as possible [8], while decreasing the number of variables. The analysis of CDP versus TEP; although, very valuable to show the process drift over time, only illustrates the change among two variables. PCA in contrary compresses the information into smaller number of variables, which can be used for early change detections among all variables.

Plotting principle component 1 versus principle component 2 by years, visualizes the maximum information among the variables over the years of dataset (Figure 4) in a two dimensional chart. As can be seen, the principle component plots indicate change in the process over years. Application of latent variables (e.g. PCA) over time is a method for discovering process change over time which can be used for early detection of degradation for preventative maintenance [8].

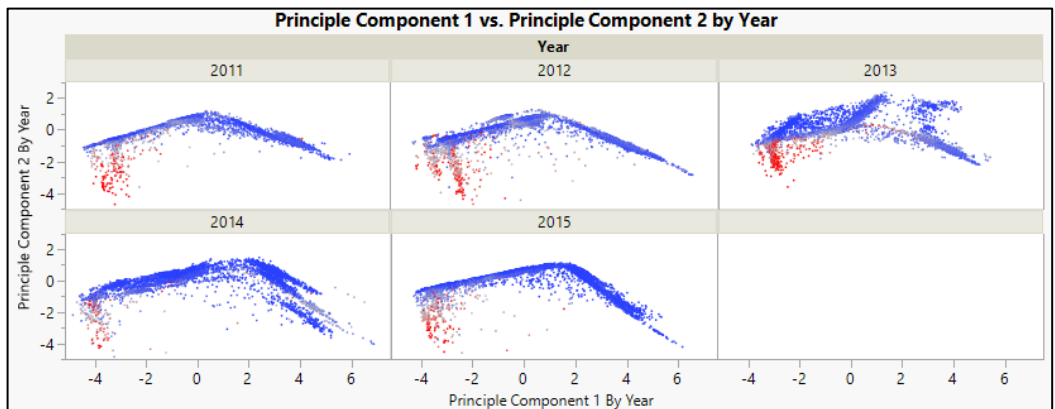

*Figure 4. Principle component 1 versus 2 by year. Red data points indicate higher values for NOx.*

## 7 NOx Prediction using K-Nearest-Neighbour Algorithm

The topic of PEMS predictive modelling have been discussed in literature by many authors for the last 25 years [17]. Most successful algorithms have been non-parametric, which do not model the process based on statistical distributions or mathematical models, simulating behavior of process under study. Instead relying on application of previous history for finding an approximation to current parameters [18].

K-Nearest-Neighbor (KNN) is a machine learning algorithm, which could be used for both classification and regression, approximating a value based on K closest training data points. KNN is considered a non-parametric algorithm, meaning the approximation is not based on any specific distribution (e.g. Poisson,

Gamma or Normal), instead utilizing training dataset and finding weighted average value of K nearest neighbors based on distance to the neighbors, identified in training dataset.

As discussed in section 6, gas turbine electrical generation process goes through subtle change among relationship between process parameters. As a result success rate of models predicting NOx production over time drops, due to effects of process change. This section compares two different models based on KNN, first model for all years together and the second model for each year analyzed separately.

Creating KNN predictive model for all years together produced the best accuracy using three neighbors, which gives the lowest root average square error (RASE). Figure 5, illustrates relationship between RASE and number of K for predicting NOx using KNN algorithm. This means by using only 3 neighbors for calculating distance averaging NOx, the lowest error is achieved.

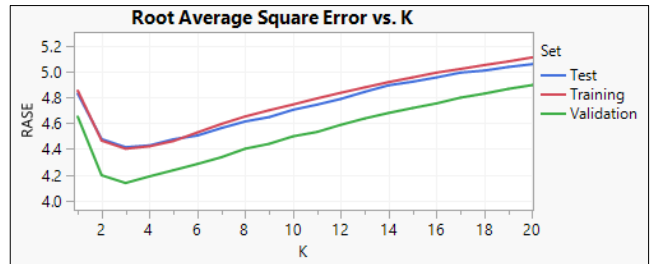

*Figure 5. Figure 5 Root average square error versus K*

KNN is a powerful algorithm for datasets with localized data point concentrations, due to reduction of average distance to each neighbor. As observed in Figure 3 and Table 7, gas turbine operations are mostly performed within a small number of operational modes which causes data points to be relatively concentrated and forming high density locales. This characteristic, increases performance of KNN for predicting process outcomes; i.e. NOx production. Meanwhile, increasing number of variables (i.e. dimensions), decreases KNN performance known as "Curse of Dimensionality", which has been discussed among academia in depth [19]. In case of high number of variables (dimensionality), KNN prediction performance usually drops; therefore, dimension reduction techniques such as principle components or similar methods are highly recommended for large number of data variables[13].

Table 8, illustrates performance comparison of KNN for all years together versus yearly generated models. As Table 8 indicates KNN performance is slightly higher using yearly models and indicators such as R-Squared, RASE (Root Mean Squared Error) and AAE (Average Absolute Error) exhibit better performance when the prediction models are based on annual parameters. This is an important observation by this paper indicating the process is changing over time and although annual models have less training data; however, produce better results.

|  | Predicted KNN NOX (mg/m3) All Years | | | Predicted KNN NOX (mg/m3) By Year | | | Freq |
| --- | --- | --- | --- | --- | --- | --- | --- |
|  | R-Square | RASE | AAE | R-Square | RASE | AAE |  |
| Training | 0.9349 | 2.9770 | 1.7471 | 0.9469 | 2.6874 | 1.5945 | 28554 |
| Validation | 0.8693 | 4.1359 | 2.5693 | 0.8919 | 3.7603 | 2.3441 | 4078 |
| Test | 0.8634 | 4.4142 | 2.6098 | 0.8934 | 3.8999 | 2.3345 | 4079 |
| Total | 0.9196 | 3.3104 | 1.9343 | 0.9348 | 2.9796 | 1.7600 | 36711 |

*Table 8. Model performance comparisons of overall data together versus year by year*

Table 9 illustrates NOx prediction performance of KNN for each given year, indicating total annual R-Squared were above 90 percent, with the lowest for 2013 (90%) and highest for 2015 (94%). This observation is indication of process characteristics change over time which using smaller more localized datasets produce higher prediction rates than using larger datasets.

| KNN By Year Details | | | | | | | | | | | | |
|---|---|---|---|---|---|---|---|---|---|---|---|---|
| | Training | | | Validation | | | Test | | | Total | | |
| Year | R - Square | RASE | AAE | R - Square | RASE | AAE | R - Square | RASE | AAE | R - Square | RASE | AAE | Freq |
| 2011 | 0.9516 | 2.3292 | 1.4023 | 0.8960 | 3.3790 | 2.1079 | 0.8940 | 3.7366 | 2.1268 | 0.9383 | 2.6531 | 1.5611 | 7411 |
| 2012 | 0.9392 | 2.5264 | 1.5945 | 0.8756 | 3.6526 | 2.4075 | 0.8881 | 3.3106 | 2.1944 | 0.9267 | 2.7686 | 1.7516 | 7628 |
| 2013 | 0.9215 | 3.3745 | 2.0861 | 0.8574 | 4.5536 | 2.9462 | 0.8373 | 4.8497 | 3.0821 | 0.9051 | 3.7115 | 2.2924 | 7152 |
| 2014 | 0.9254 | 2.6972 | 1.4452 | 0.8321 | 3.6502 | 2.1397 | 0.8393 | 4.3880 | 2.2232 | 0.9054 | 3.0456 | 1.6086 | 7136 |
| 2015 | 0.9539 | 2.4099 | 1.4557 | 0.8955 | 3.4772 | 2.1291 | 0.9244 | 2.9827 | 2.0707 | 0.9447 | 2.6170 | 1.5988 | 7384 |

*Table 9. Comparison of predictive model performance by year*

Analysis of residuals (Figure 6) illustrate consistently scattered residuals across the plot without any specific pattern, which is a positive sign of strong KNN prediction model, independent of any significant bias. The data points were broken down for each year to Training (70%), Validation (15%) and Test (15%). The exact number of the break downs can be seen under column Frequency in Table 8 and 9.

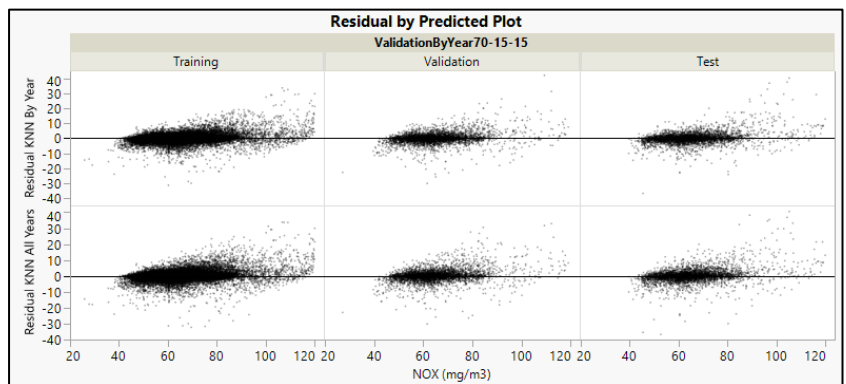

*Figure 6. Analysis of residuals*

Figure 7 displays the plot of actual NOx emissions versus predicted over time. As can be seen the predictions are very close to actual, which is also supported by high R-Squared values of above 90%. KNN, a non-parametric algorithm provides high success rates for NOx prediction, by finding the previous K similar observations at the training data and then approximating the new value based on the distance from each observation.

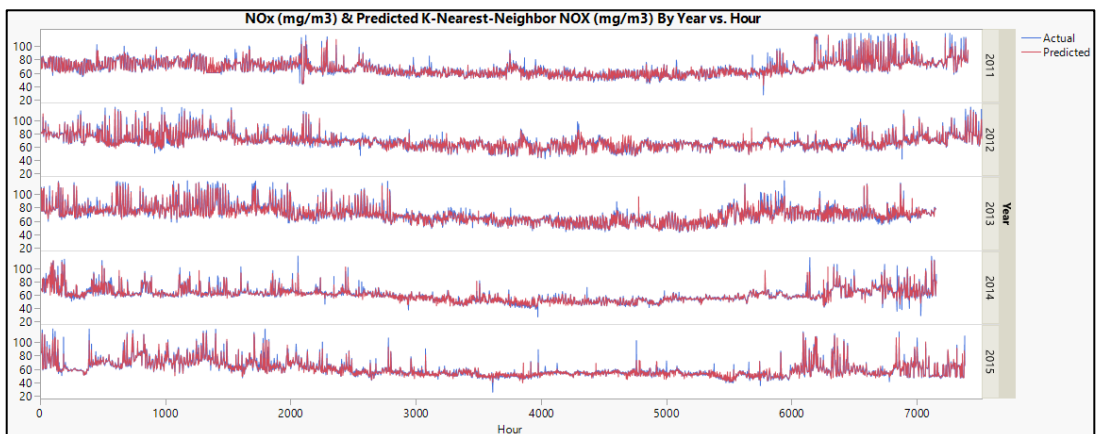

*Figure 7. NOx K-Nearest-Neighbor prediction vs. actual*

# 8 Discussions

## 8.1 Process Change

The process of power generation is dependent on many dynamic factors including power demand, weather, equipment efficiencies and operational conditions. Therefore, predicting NOx which is a process dependent pollutant can be more effectively accomplished using shorter training datasets which are more similar to current operating parameters. Hence, adaptive algorithms may offer more advantage since they assign heavier weight to more recent training data.

## 8.2 Digital transformation and open data challenge

Industrial data analytics can only be accomplished upon availability of accurate datasets describing a dynamic process. Applications of machine learning and predictive modelling will further advance by availability of data for finding the most effective methodologies, openly discussing the results in dissertations, conferences and scientific publications. Sharing non-confidential industrial data creates the opportunity for a larger community of research and training enthusiasts, creating next generation of digital savvy workers, benefitting the very same industries by accelerating adoption of new technologies.

# 9 Future Work

Throughout the paper there were discussions of process degradation. However, degradation analysis needs to be quantitatively defined, monitored and if possible minimized. Formulation of degrading process requires more detailed parameters than were available in this dataset. For instance Brayton cycle thermodynamics of gas turbine parameters could be used for monitoring the gas turbine performance, predicting degradations and efficiencies.

Adaptive KNN algorithm in pollutant release prediction as well as other applications that have a changing behavior can be explored. Modelling a dynamic process requires a resilient adaptive algorithm to incorporate gradual change as it might be due to social and economics evolution or an industrial equipment degradation.

# 10 Acknowledgement

Wayne Hovdestad, M.Eng., P.Eng., has kindly collaborated with the research on the engineering analysis of the gas turbine parameters. His insight added extreme values to better understanding the gas turbine thermodynamics and detailed operations.

The data used in this analysis was produced by a power generation plant in north western Turkey which is a well populated and industrialized section of the country. This research could not be possible without the shared data with the scientific community of AI/ML. Author would like to express gratitude to the power generator operator for publishing and sharing the operational data. My best gratitude to Dr. Heysem Kaya, for facilitating this dataset to be publicly available and also collaborating with the author for this research.

Author would like to acknowledge CFREF (First Canadian Research Excellence Fund) for providing the opportunity to research new topics helping environment protection as well as academia.

Extreme gratitude to SAS Institute for design and implementation of JMP, which enabled the author to perform complex analytics on short time and finding the most effective methods of data mining and prediction modelling.